# Accurate non-iterative phase retrieval for time-resolved coherent diffraction imaging at X-ray free electron lasers

Sung Yun Lee[1,2,3], Eunyoung Park[1,2,3], Sinwoo Kim[1,2,3], Jungchan Choi[1,2,3], Seung-Phil Heo[1,2,3], Junha Hwang[2,4], Do Geun Jang[4], Daewoong Nam[2,4], Sangsoo Kim[4], and Changyong Song[1,2,3,*]

*[1]Department of Physics, POSTECH, Pohang 37673, Republic of Korea; [2]Photon Science Center, POSTECH, Pohang 37673, Republic of Korea; [3]Center for Ultrafast Science on Quantum Matter, Max Planck POSTECH/Korea Research Initiative, Pohang 37673, Republic of Korea; [4]XFEL Beamline Division, Pohang Accelerator Laboratory, POSTECH, Pohang 37673, Republic of Korea; [*]Corresponding author, e-mail: cysong@postech.ac.kr*

**Time-resolved coherent diffraction imaging (CDI) with ultrashort, ultraintense X-ray pulses from X-ray free-electron lasers (XFELs) has enabled the ultrafast tracking of non-equilibrium nanoscale dynamics in various materials, opening a new horizon in ultrafast science. However, current studies rely heavily on extensive human effort to retrieve the phases of single-pulse diffraction patterns, a critical bottleneck as research is accelerated by high-repetition-rate XFELs and large language models. Here we introduce k2rNN, a neural network that reconstructs *r*-space images from *k*-space patterns in a single step, together with Diffraction model V2, a physics model that generates realistic patterns. On simulated data and on experimental data from one 12-hour shift at PAL-XFEL, k2rNN outperforms existing algorithms and models in accuracy and robustness, and its lightweight variant reaches a throughput of 1,240 patterns per second. Together, they offer a significant methodological advance in time-resolved CDI, facilitating next-generation ultrafast studies.**

## INTRODUCTION

Since the advent of X-ray free-electron lasers (XFELs) capable of generating ultrashort, ultraintense X-ray pulses, ultrafast X-ray science has advanced rapidly, leading to numerous important scientific discoveries[1]. A single such pulse can capture the diffraction pattern of a nanoparticle before the particle is destroyed, so that its structure at a given moment is recorded clearly, which makes ultrafast measurements practical[2,3]. In this setting, coherent diffraction imaging (CDI) enables lensless imaging of individual nanoparticles by reconstructing real-space ($r$-space) images, maps of the projected electron density, from reciprocal-space ($k$-space) diffraction patterns whose phases the detector does not record[4–6]. Combined with an optical pump, time-resolved CDI has become a powerful microscope for non-equilibrium nanoscale dynamics in various materials, following the laser ablation of nanostructures[7], the superheating of clusters[8], the picosecond melting and disintegration of metallic nanoparticles[9,10], non-equilibrium surface atomic ordering[11], and the energy transfer and plasmon-controlled deformation of gold nanoparticles[12,13]. Because every pulse probes a fresh particle, the dynamics are traced statistically from thousands of single-particle snapshots at each delay; in this way, time-resolved CDI has opened a new horizon in ultrafast science.

This strength comes at a price: every snapshot must be reconstructed individually, and in current practice this relies heavily on human effort. Phase retrieval from a single-pulse pattern is difficult, because the $r$-space support is unknown and the pattern is noisy, partly hidden by the beamstop and detector gaps, and contaminated by contributions that the algorithms do not account for[14,15]. Iterative algorithms such as hybrid input–output (HIO)[16] and generalized proximal smoothing (GPS)[17], combined with shrink-wrap[18], therefore need hundreds to thousands of iterations from many random initial phases, and good images still depend on an expert who tunes an appropriate $r$-space support and the algorithm's parameters and inspects every result. Even so, the algorithms often stagnate on weak or contaminated patterns, which are then discarded. As a result, the measured patterns are first sorted by human inspection, and images are then carefully reconstructed from the selected patterns one by one, a process that can take weeks to months. A single 12-h shift at PAL-XFEL, analysed here, yielded 13,388 hits, and a typical beamtime of six shifts yields up to about 100,000.

The bottleneck will become more severe as ultrafast science accelerates. Superconducting XFELs deliver up to 27,000 pulses per second at the European XFEL[19] and up to a million at LCLS-II[20], and single-particle imaging records millions of patterns per day[21]. Hit finding already runs on-line[22], but phase retrieval does not, so experimenters cannot see the objects while they measure and can only guess their shapes from the patterns. At the same time, agents built on large language models have begun to plan and run experiments[23] and to operate X-ray beamlines[24,25]. To work with time-resolved CDI, such agents need phase retrieval that is fast and free of human supervision, so that they can reason about objects rather than raw patterns.

Deep neural networks offer a way out. Besides denoising weak patterns for iterative phase retrieval[26], networks can reconstruct an image from a pattern in a single forward pass, fast enough for real-time imaging[27–30]. In real experiments, however, a network is only as accurate as the physics model behind its training patterns, and most networks are designed for a specific class of samples, such as strained nanocrystals[31] or biological macromolecules[29]. The deep phase retrieval (DPR)[30] instead learns from general objects, but its diffraction model omits factors present in measured patterns: strong phase shift and absorption by dense, thick particles, Ewald-sphere curvature at high scattering angles[14,32], and contamination by scattering from the sample environment or apertures, or by detector miscalibration[33,34].

Here we close this gap on both fronts, the training data and the network (Fig. 1a). Diffraction model V2 is a physics model that generates patterns close to those measured in real experiments by including Ewald-sphere curvature, a complex refractive index without the weak-scattering approximation and a broad range of measurement effects absent from DPR's model. k2rNN, trained on these patterns, reconstructs an *r*-space image from a *k*-space pattern in a single step: a vision foundation model (VFM)[35] reads the pattern, a vision transformer on *k*-space tokens (*k*-space ViT)[36] predicts the full complex field, and after an inverse Fourier transform, cascaded fast Fourier convolutions (*r*-space FFC refiner)[37] refine the image while keeping it consistent with the measured amplitudes. The network thus completes the measured pattern and recovers its phases without iterative optimization. On simulated patterns, its peak signal-to-noise ratio (PSNR) exceeds those of conventional iterative algorithms and of DPR by about 15 dB, even when DPR is retrained on our patterns. On more than 1,000 gold-nanorod patterns from a single 12-h pump–probe shift at PAL-XFEL, its reconstructions agree best with the measurements and allow a general-purpose vision encoder to sort the entire shift into groups that streamline human inspection. It also works on other facilities and samples without retraining, and its lightweight version reconstructs 1,240 patterns per second, bringing on-line phase retrieval within reach.

## RESULTS

### ***Diffraction model V2***

Diffraction model V2 generates every training pattern from a new random object and new experimental conditions (Fig. 1a,b; Methods; Supplementary Table 1). As in DPR[30], the object is built from a shape (an EMNIST character[38]) and a density (a CIFAR-100 image[39]), but V2 treats it as a physical particle with a complex refractive index $n = 1 - \delta + i\beta$. Its exit wave, $\exp[(i - \beta/\delta)\varphi] - 1$, is used in its full exponential form, without weak-phase or weak-absorption approximations; the network returns $\varphi$, which is proportional to the projected electron density. The far field of this wave is then computed on a curved Ewald sphere. The projection approximation neglects the component of the scattering vector $\boldsymbol{q}$ along the beam, but at a scattering angle $2\theta$ this component, $q_z \approx |\boldsymbol{q}_\perp|^2/(2k)$, becomes significant, so that different depths of the particle scatter with different phases. We expand this phase in powers of the squared scattering angle and weight each term with the corresponding moment of the particle's depth profile (Methods). With random curvature strengths and mid-plane positions, the training patterns range from the thin limit, where the series reduces to the projection, to thick particles whose patterns lose Friedel symmetry and show curvature-dependent fringes[14,32]. Strong phase and absorption and Ewald-sphere curvature are the main departures from DPR's model, and both change the pattern in ways that a network trained on projections would misread as object structure.

The remaining terms reproduce real measurement conditions. The beam is partially coherent, following a Gaussian Schell model[40]. A radial background is added to 20% of the patterns, and streaks through the beam centre, mimicking stray scattering from apertures and slits[33], to 10%. The detector model adds noise, gain and offset errors, bad pixels[34], imperfect background subtraction and beam-centre errors. Masks are then applied to the simulated patterns, combining irregular regions from the NVIDIA irregular-mask dataset[41], a beamstop and, in half of the patterns, detector-module gaps.

### *k2rNN*

VFMs pretrained on natural images are known to transfer well to specialized tasks, from segmenting microscopy images[42] to estimating depth, which a single photograph does not record[43]. k2rNN likewise builds on a VFM to recover the phases that a diffraction pattern does not record; the VFM extracts features from the pattern for *k*-space tokens, and two subsequent modules reconstruct the image (Fig. 1a, Supplementary Fig. 1 and Supplementary Table 2). For the VFM, the pattern $\boldsymbol{I}$ and its valid-pixel mask $\boldsymbol{M}$ are encoded as an image with normalized $\ln(1 + \boldsymbol{I})$, $\boldsymbol{M}$ and $\boldsymbol{I}$ in its RGB channels (Supplementary Fig. 2). We use DINOv3 ConvNeXt-B[35] as the VFM backbone and fine-tune it at a reduced learning rate.

The first module, the *k*-space ViT, works directly in *k*-space, like a recent unsupervised Fourier ViT for Bragg CDI[44]. The 512 × 512 pattern is divided into 4,096 tokens of 8 × 8 pixels, each carrying its log intensity, mask and position $\boldsymbol{q}$ together with its VFM features. Transformer blocks combine attention between neighbouring tokens[36] with register tokens that gather information from the whole pattern. Each token predicts the log amplitude and phasor of a 2 × 2 patch in the 128 × 128 complex field, which exactly samples the far field of the 64 × 64 *r*-space window. The predicted amplitudes are blended with the measured ones by a learned, pixel-wise weight, and an exact inverse Fourier transform yields an initial *r*-space image. The second module, the *r*-space FFC refiner, corrects this image in four stages of three residual FFC blocks[37,45] and, between stages, blends its *k*-space amplitudes with the measured ones to keep the refinement consistent with the data.

### *Performance on simulated data*

We trained k2rNN for 320,000 steps, each on 160 freshly simulated patterns, a total of 51.2 million patterns (Methods). In separate runs with increasing numbers of training steps, its median PSNR rises steeply up to 40,000 steps and then saturates, gaining only 0.1 dB between 160,000 and 320,000 steps (Fig. 2a). During training, its mean validation PSNR reaches 36.7 dB, whereas a non-ViT variant of k2rNN peaks at 35.8 dB after 200,000 steps and then declines, supporting the suitability of our *k*-space ViT (Fig. 2b). DPR retrained on Diffraction model V2 data peaks far lower, at 18.9 dB with DPR's own loss function (DPR+) and 21.8 dB with ours (DPR++). On one NVIDIA RTX 6000 Ada GPU, k2rNN processes 134 patterns per second, close to DPR's 172 and more than 1,000 times faster than GPS (Fig. 2c).

On the test patterns, k2rNN reaches a median PSNR of 35.3 dB and a median structural similarity (SSIM) of 0.986, about 15 dB and 0.19 higher than the best of the conventional algorithms and DPR variants (Fig. 2d,e). Because measured data have no ground truth, we also scored each reconstruction by how well it explains its pattern, using two metrics averaged over shells of constant $|\boldsymbol{q}|$: the *k*-space *R*-factor $\langle R_{\mathrm{F}}\rangle_q$ (a normalized L1 error of the amplitudes) and the Poisson deviance $\langle D\rangle_q$. Both compare the pattern with a forward model of the reconstruction, matching its total counts and fitting its coherence length (Methods). k2rNN's median values (0.211 and 1.26) are close to those of the true objects (0.207 and 1.22) and lower than those of every other method (Fig. 2f,g). Among the other methods, GPS tracks the patterns most closely (0.247 and 1.79), yet its objects have the lowest PSNR: from incomplete, contaminated patterns, it returns false objects that explain them almost equally well. Data consistency alone therefore does not guarantee accuracy, so reconstructions of measured data still need careful inspection. In the examples, the conventional algorithms miss the internal density and the DPR variants distort the outline, whereas k2rNN recovers both (Fig. 2h).

### *Application to a real pump–probe experiment*

We applied k2rNN to single-shot patterns of gold nanorods (145 nm in length and 50 nm in diameter), measured at the coherent X-ray imaging endstation (NCI-CXI) of PAL-XFEL[46,47] before and up to 150 ps after excitation by femtosecond 400-nm pump pulses, during the study that reported their plasmon-controlled deformation[13] (Methods). These data, 30,457 shots or 511 GB of JUNGFRAU 4M detector frames, were recorded in one 12-h shift. An automated pipeline found 13,388 hits among the 30,245 usable shots, and k2rNN reconstructed all of them in 219 s on four NVIDIA RTX 6000 Ada GPUs, including scoring the reconstructions and extracting their embeddings; the whole pipeline took about an hour, nearly 90% of it for reading the detector frames and the compressed caches over the network. All its steps, from the detector frames to clustering, run in CDI Workbench, a graphical application we developed for this work (Methods).

Because k2rNN returns *r*-space images rather than *k*-space patterns, the whole shift can be sorted with general-purpose image tools: each reconstruction is 64 times smaller than its pattern and shows the particle in a form that both people and vision encoders pretrained on natural images conveniently read. We embedded every reconstruction with the vision encoder of a locally running vision–language model, Qwen3.5-9B[48], and clustered the embeddings into nine groups (Fig. 3a;

Methods). Streak-dominated groups and hits without excess photons or with empty reconstructions were set aside automatically; after viewing the representatives of the remaining groups, we kept five (6,717 hits) and set aside two that held patterns too weak or contaminated for analysis. We then labelled 1,080 hits by hand, guided by automatic recommendations, as single (710) or multiple (370) particles, all but two of them from the kept groups. Two groups match these labels closely: group #2 holds 82% of the singles with a label purity of 98.6%, and group #6 holds 87% of the multi-particle hits with a label purity of 99.7%. From detector frames to labelled data, people thus only had to look at *r*-space images of grouped hits, never at diffraction patterns, which made inspection much more efficient.

On single rods, k2rNN fits the data better than every other method in both metrics (Fig. 3b,c). For intact rods (no pump, or pump–probe delay $\Delta t \leq 0$), its median $\langle R_F \rangle_q$ is 0.158, compared with 0.168 for DPR++ and 0.161 for GPS, and its median $\langle D \rangle_q$ is 0.68, compared with 0.70 and 0.77. For pumped rods ($0 < \Delta t \leq 70$ ps), the margins widen: $\langle R_F \rangle_q$ 0.170 versus 0.184 and 0.176; $\langle D \rangle_q$ 0.85 versus 0.87 and 0.98. Resolved by delay, k2rNN's $\langle D \rangle_q$ is lower than that of GPS at every delay, and its $\langle R_F \rangle_q$ for intact rods and at 5–40 ps (Fig. 3d,e). In the examples, all methods recover the intact rod (Fig. 3f). However, as the rod deforms, DPR++ blurs it, and HIO and GPS return rods with mottled, noisy interiors (50 ps) before failing on the most deformed rod (70 ps), whereas k2rNN returns a compact particle with a well-defined outline at every delay shown.

***Deformation of single nanorods across delays***

Because every hit is reconstructed, the rods can be analysed statistically. We measured each rod's aspect ratio, orientation and support area and classified the rods as round or, by orientation, as horizontal, diagonal or vertical (Fig. 4a,b; Methods). The unpumped rods have a median aspect ratio of 3.0, close to the nominal 2.9, and a median support area of 6,100 $nm^2$. The shares of the three orientation classes are similar at all delays, as expected for randomly oriented rods (Fig. 4c).

One shift yields only tens of rods per delay and class, but the pooled statistics show a clear deformation trend, although resolving the kinetics will require more shots (Fig. 4d,e). The rods keep their intact aspect ratio and area up to 1 ps, and from 5 ps onwards, the aspect ratio falls and the apparent support area grows, reaching 2.0–2.5 and 7,400–8,800 $nm^2$ at 20–70 ps. The rods thus become shorter and wider, consistent with the ultrafast deformation of photoexcited gold nanorods reported previously[13]. Horizontal rods, parallel to the pump polarization, swell first (larger area than diagonal and vertical rods at 10 ps), and vertical rods keep their aspect ratio longest (above diagonal rods at 20 ps), as in the previous report[13]. The delay tree also shows this sequence in representative reconstructions (Fig. 4f).

### *Performance on multiple particles*

Hits with several particles test the method further, because pairs and clusters of rods form a larger, less compact object without a simple support. On the 369 multi-particle hits up to 70 ps, k2rNN again fits the data better than every other method in both metrics (Fig. 5a,b). Compared with GPS, the best of the other methods, its median $\langle R_F \rangle_q$ is 0.171 versus 0.173 for intact particles and 0.188 versus 0.209 for pumped ones, and its median $\langle D \rangle_q$ is 1.07 versus 1.10 and 1.12 versus 1.51. In the examples, only k2rNN reconstructs every hit cleanly: it separates side-by-side, crossed and clustered rods and preserves their shapes, whereas DPR++ distorts them, and HIO and GPS return striped interiors or break the crossed rods into spurious fragments (Fig. 5c).

### *Robustness to contamination and generality across experiments*

Two measured nanorod patterns illustrate how contamination affects each method (Supplementary Fig. 3a). From a weak-signal pattern, k2rNN and DPR++ return the rods even before background subtraction, whereas the conventional algorithms fail. On a pattern crossed by a streak, k2rNN returns nearly the same particle whether the streak is masked or not, whereas the other methods need masking, and HIO fails even with it. In both patterns, the registers of the *k*-space ViT concentrate on the informative speckles (Supplementary Fig. 3b).

Without retraining, k2rNN also reconstructs patterns from other facilities, detectors and samples (Supplementary Fig. 4). For a single and a double silver flower recorded with a multi-port CCD detector at PAL-XFEL, each missing a quarter of the pattern behind a beamstop[30], it returns both particles with the lowest $\langle D \rangle_q$ of all methods. For 310 pnCCD patterns of the viruses PBCV-1 and T4, recorded at LCLS and deposited in the Coherent X-ray Imaging Data Bank[49,50] (Methods), its median $\langle R_{\mathrm{F}} \rangle_q$ and $\langle D \rangle_q$ are the lowest of all methods for both viruses; the differences are significant against DPR and HIO in both metrics, and in $\langle D \rangle_q$ against DPR++ for both viruses and GPS for T4.

### *Lightweight models and real-time deployment*

For real-time use at the instrument, we built k2rNN-S, a lightweight variant with the same design but a StarNet backbone[51], a k-space ViT over 1,024 tokens and a narrower, deeper refiner (Methods). With StarNet S1, k2rNN-S has 6.9 million parameters and needs 24 GFLOPs per pattern, compared with 102 million and 302 GFLOPs for k2rNN, yet its mean test PSNR of 33.7 dB is only 0.8 dB lower; with StarNet S4, it reaches 34.3 dB (Fig. 6a,b).

For deployment, we converted each network into a TensorRT FP16 engine with custom Fourier-transform kernels (optimized FP16 engine; Methods). The engines keep the accuracy of the original models (mean PSNR within 0.07 dB) and process 210 patterns per second for k2rNN and 1,240 for k2rNN-S S1 (Fig. 6b), the latter a third of the 3,520 frames per second that the AGIPD detector records at the high-repetition-rate European XFEL[52]; a few parallel engines could thus keep pace with it. On the measured hits, the two engines do not differ significantly in $\langle R_{\mathrm{F}} \rangle_q$, and k2rNN's $\langle D \rangle_q$ is only slightly lower for pumped and multiple particles (Fig. 6c,d).

**DISCUSSION**

A single forward pass of k2rNN reconstructs single-pulse diffraction patterns at least as consistently with the data as the best iterative algorithm, and more consistently on pumped and multiple particles, where iterative phase retrieval struggles, while running about a thousand times faster without per-pattern tuning. Two components make this possible. First, Diffraction model V2 reproduces the effects that make measured patterns depart from the projection model: retrained with our training scheme (DPR++), even DPR fits the measured rods better than its original version. Second, the architecture predicts the complex field directly in k-space and refines the r-space image with data consistency, outperforming every other method tested, including DPR++, on both simulated and measured patterns (Figs. 2d, 3b,c and 5a,b).

This changes how time-resolved CDI experiments are analysed. Once every hit is reconstructed, here the 13,388 hits of a 12-h shift in a few minutes, the reconstructions rather than the patterns become the data: people inspect images directly, without retrieving phases themselves, and ultrafast dynamics can be investigated without discarding patterns on which iterative algorithms fail. With k2rNN-S, a pattern takes less than 1 ms of GPU time, so the bottleneck moves to data handling, which took nearly 90% of our pipeline's hour; at future high-repetition-rate XFELs, on-line reconstruction will need pipelines that immediately stream frames from the detector to the GPUs. Reconstruction at the detector could then select informative snapshots and steer the delay scan, pump fluence or sample delivery while the experiment runs.

Reliable phase retrieval also opens new experiments. Snapshots of dimers, chains and clusters of plasmonic particles could show how near-field coupling[53] drives ultrafast reshaping or sintering, which single-particle experiments cannot access. Trained only on single objects, k2rNN already fits multi-particle patterns better than GPS, and training on multi-particle objects should widen this lead. Reconstructed images are also natural input for vision–language models: the local model in CDI Workbench here only named the groups and summarized the results, but such models and agents that operate X-ray beamlines[24,25] could comment on the data, flag new structures and propose the next measurement.

Our approach has several limitations. The network reconstructs a real, non-negative projected phase in a 64 × 64 window from a 512 × 512 pattern; recovering the full complex refractive index, larger fields of view or other oversampling ratios require retraining or binning. Diffraction model V2 omits the illumination profile, wide-angle geometry and multiple scattering. The fitted metrics test data consistency, not uniqueness, and resolution estimates such as the phase retrieval transfer function rely on many reconstructions from random initial phases[14], so single-shot estimates remain open. Within these limits, k2rNN turns phase retrieval from the slowest and most labour-intensive step of a time-resolved CDI experiment into a real-time, automated one.

## METHODS

### *Diffraction model V2*

Each pattern is generated on the GPU from a new random object and new measurement conditions in five stages: object, far field, intensity, detector and mask. The parameter ranges are set wide enough to cover the measured data rather than fitted to one instrument; they are listed in Supplementary Table 1.

*Object*. A binarized EMNIST character[38], randomly flipped, rotated and scaled to a longest side of 20–60 pixels, is multiplied by a greyscale CIFAR-100 image[39] to give the normalized projected thickness $s$; the object is $sD$ thick along the beam for a maximum thickness $D$. The object has a complex refractive index $n = 1 - \delta + i\beta$, so its exit wave relative to the incident wave is $o = \exp[(i - r)\varphi_{\max}s] - 1$, with $r = \beta/\delta$ and the maximum phase $\varphi_{\max}$ drawn randomly. The training target is the projected phase $\varphi = \varphi_{\max}s$.

*Far field*. Ewald-sphere curvature gives the scattering vector a longitudinal component $q_z \approx |\boldsymbol{q}_\perp|^2/(2k)$, which makes the pattern depend on the depth within the object. In the column approximation, with the scattering of each column spread uniformly along its depth, expanding the phase factor $\exp(-iq_z z)$ at depth $z$ in a power series gives (Supplementary Note 1)

$$F(\boldsymbol{q}_\perp) = \sum_{n=0}^{\infty} \frac{(-i\kappa\rho^2)^n}{n!} \mathrm{FT}[om_nR](\boldsymbol{q}_\perp),$$

where $\rho$ is the normalized distance from the beam centre, $\kappa$, the largest value of $q_zD$, sets the curvature strength, $m_n$ is the column average of the $n$-th power of the normalized depth, and the phase ramp $R$ shifts the beam centre by up to 2 pixels. The series expands the shape transform $\exp(-i\kappa\rho^2\zeta_0)$ $\mathrm{sinc}(\kappa\rho^2 s/2)$ of a column centred at depth $\zeta_0$ into terms that each need only one Fourier transform, and it reduces to the projection approximation used by DPR for $\kappa \to 0$. Through these terms, the pattern of a thick object is no longer centrosymmetric even for a real exit wave, and its high-angle speckles are shifted and damped; a network trained on simple projections would read these departures as object structure.

*Intensity, detector and mask.* The intensity $|\boldsymbol{F}|^2$ is blurred by the partial-coherence kernel of the Gaussian Schell model[40] and scaled to $10^5$–$10^7$ photons. Radial backgrounds and streaks are added to 20% and 10% of patterns, respectively, and the detector applies Poisson noise, gain errors, readout noise, offsets and bad pixels. The background is then subtracted with an error of up to ±20%, leaving residual background of either sign. Irregular masks[41], a beamstop and detector-gap bands leave on average 24.5% of the pixels unmeasured.

***Network architecture***

The architecture is shown in Supplementary Fig. 1, and its hyperparameters are listed in Supplementary Table 2. The background-subtracted pattern $\boldsymbol{I}$ and its valid-pixel mask $\boldsymbol{M}$ are mapped to an RGB image, $R = \ln(1 + \boldsymbol{I})/\max(\cdot)$, $G = \boldsymbol{M}$ and $B = \boldsymbol{I}/\max(\cdot)$ (Supplementary Fig. 2), with two scalars describing the total photon count and the valid-pixel fraction. The image is encoded by the pretrained DINOv3 ConvNeXt-B[35], and the feature maps of its four stages are resized to 32 × 32 and fused into one.

In the $k$-space ViT, each 8 × 8 patch of the pattern forms a token from its pixel values, the magnitude and Fourier features[54] of $\boldsymbol{q}$ and the VFM features. The 4,096 tokens are ordered along a Z-order curve as in serialized point transformers[36] and pass through eight blocks of self-attention within overlapping windows of 256 tokens, combined with cross-attention to and from 16 register tokens. Each token predicts the log amplitude and phasor of a 2 × 2 patch of the 128 × 128 Fourier grid, which samples the far field of the 64 × 64 object window at an oversampling ratio of 2 and thus without loss.

The $k$-space mixer blends a predicted Fourier field $\boldsymbol{Y}$ with the measured amplitudes: a pixel-wise MLP gives a weight $\boldsymbol{\gamma} \geq 0$, set to zero at unmeasured pixels, the amplitude $(|Y| + \gamma\sqrt{I_{\downarrow 4}})/(1 + \gamma)$ is combined with the predicted phase, where $\boldsymbol{I}_{\downarrow 4}$ is the pattern sampled on the 128 × 128 grid, and an inverse FFT gives the *r*-space object. The *r*-space refiner corrects this object in four stages of three FFC blocks[37,45] and passes the updated object through the *k*-space mixer between stages. The output is cropped to the 64 × 64 window, clipped at zero and scaled so that its far field carries the measured photon count. k2rNN-S uses a StarNet S1 or S4 backbone[51] pretrained on ImageNet-1k, a *k*-space ViT of width 128 over 1,024 tokens of 16 × 16 pixels and an FFC refiner of width 32 with six stages.

***Training***

The loss combines L1 losses on the initial and refined *r*-space images, an amplitude-weighted cosine loss on the Fourier phases and an L1 loss on the log amplitude with coefficients of 0.5, 1.0, 1.0 and 0.1, respectively; the *r*-space terms are computed after alignment for translation and the twin image. k2rNN was trained with the AdamW optimizer, a learning rate of $5 \times 10^{-4}$ (×0.2 for the VFM backbone) with linear warm-up and cosine decay, and an exponential moving average of the weights. Each of the 320,000 training steps used 160 new patterns from Diffraction model V2 on 16 GPUs, in BF16 mixed precision except for the Fourier transforms and amplitude blending, which run in FP32. The checkpoint with the highest mean PSNR on 500 fixed validation patterns was kept. k2rNN-S was trained in the same way with synchronized batch normalization.

The non-ViT variant of k2rNN replaces the *k*-space ViT with a convolutional decoder and was otherwise trained in the same way. DPR+ and DPR++ are DPR[30] retrained on Diffraction model V2 in the same way as k2rNN, but in FP32 with synchronized batch normalization; DPR+ uses DPR's own loss, and DPR++ uses ours (learning rate of $2 \times 10^{-4}$, as training diverged at $5 \times 10^{-4}$). DPR alone denotes the published network with its pretrained weights[30].

***Iterative phase retrieval***

HIO[16] and GPS-R[17] were run with shrink-wrap[18] using the settings of DPR's study[30], 1,000 iterations from each of 100 random initial phases, and the reconstruction with the lowest Fourier *R*-factor ($R_F$) was kept. The initial support was a central 60 × 60-pixel square; for measured data, it was intersected with the region where the autocorrelation of the pattern exceeds 4% of its maximum, the initial support of the original shrink-wrap algorithm[18]. Neither algorithm was tuned for individual patterns, as in an automated pipeline.

***Evaluation metrics***

PSNR and SSIM were computed after aligning simulated reconstructions with the ground truth for translation and the twin image and scaling them to it by least squares. For the data-consistency metrics, a reconstruction $\boldsymbol{x}$ gives the model intensity $\boldsymbol{\lambda} = \eta \boldsymbol{K}_\sigma \circledast |\mathrm{FT}[\boldsymbol{x}]|^2 + \boldsymbol{b}$, with the Gaussian Schell-model coherence kernel $\boldsymbol{K}_\sigma$, the known background $\boldsymbol{b}$ (zero for simulated patterns; the background estimate for measured ones), a scale $\eta$ that matches the total counts and a coherence length $\sigma$ fitted for each pattern in the same way for every method. For the measured count $I$ of each pixel, $\langle D \rangle_q$ is the mean over 32 resolution shells, 8 pixels wide, of the shell-averaged Poisson deviance $2[I \ln(I/\lambda) - (I - \lambda)]$, and $\langle R_\mathrm{F} \rangle_q$ is the mean over the same shells of the $R_\mathrm{F}$ of $\boldsymbol{\lambda}$, using pixels with an expected count of at least 3 photons. Lower values indicate a better fit; for a perfect model, $\langle D \rangle_q$ is about 1 in shells with more than about 0.5 photons per pixel. The pixel-wise $R_\mathrm{F}$ is the $R_\mathrm{F}$ of $|\mathrm{FT}[\boldsymbol{x}]|^2$, scaled to the measured pattern by least squares, evaluated at each pixel.

***Pump–probe experiment on gold nanorods***

The time-resolved CDI experiment was conducted in one 12-h shift at the coherent X-ray imaging (CXI) endstation of the nanocrystallography and coherent imaging (NCI) beamline of PAL-XFEL[46,47]. Gold nanorods (50 nm in diameter, 145 nm in length) spin-coated onto $SiN_x$ membranes were scanned so that each shot hit fresh particles. The rods were excited by 400-nm pump pulses (pulse duration of 40 fs, fluence of 300 mJ cm$^{-2}$) in horizontal polarization and probed at delays from −1 to 150 ps (up to 70 ps analysed) by self-seeded 9-keV XFEL pulses (pulse duration of about 40 fs) focused to 1.9 μm (horizontal) × 2.5 μm (vertical) by Kirkpatrick–Baez mirrors. The patterns were recorded by a JUNGFRAU 4M detector (pixel size of 75 × 75 μm$^2$) at 3.13 m downstream of the sample, and the wide-angle frames recorded simultaneously by the JUNGFRAU 5M detector were not used. After 2 × 2 binning and cropping to 512 × 512 pixels around the beam centre, the pixel resolution was 5.61 nm.

***Data processing and analysis***

The detector frames were processed in seven steps of CDI Workbench (see Code availability): (1) indexing the 30,457 shots by delay and pump setting; (2) determining the beam centre and frame statistics; (3) preprocessing the frames (offset correction, masking, module assembly, cropping and binning) and building compressed caches for fast reading (511 GB → 11 GB); (4) fitting the global gain for photon conversion; (5) estimating the backgrounds and finding hits from the shot intensities (13,388 hits among 30,245 usable shots); (6) reconstructing the hits by k2rNN and computing their metrics and their embeddings by the vision encoder of Qwen3.5-9B[48]; and (7) grouping the 12,235 non-empty reconstructions of hits with excess photons by $k$-means clustering of their embeddings ($k$ = 9, automatically chosen by clustering stability under subsampling). The pipeline took about 67 min, 57 min of it for step 3, which reads the detector frames and writes the caches over the network, and 1–2 min for reading the caches.

CDI Workbench also ranks unlabelled hits for guiding inspection by a logistic-regression classifier trained on the embeddings of the hits labelled so far, conducts basic shape and delay-tree analyses and lets the local vision–language model Qwen3.5-9B propose a name and description for each group from its representative reconstructions. The two groups above the largest gap in the group medians of $\langle R_\mathrm{F} \rangle_q$ were set aside automatically, and after checking the representatives of the other seven, we kept five groups (6,717 hits). From their $r$-space images, 1,080 hits, all but two from the kept groups, were labelled by hand as single (710) or multiple (370) particles, and the 1,063 at delays up to 70 ps were analysed. Every method was given the same background-subtracted patterns and masks.

For the shape analysis, the support of each single-particle reconstruction was estimated as its connected component carrying the most intensity above Li's minimum cross-entropy threshold[55], with holes filled, and its area, aspect ratio and long-axis angle from the horizontal were computed from its second moments. Particles with an aspect ratio below 1.1 were classified as round and the others by their long axis as horizontal (0–22.5°), diagonal (22.5–67.5°) or vertical (67.5–90°). Within each class and delay, the reconstructions were grouped by aspect ratio into the nodes of the delay tree, and the flows between the nodes of consecutive delays were computed by unbalanced optimal transport[56] on the cosine distances between their mean embeddings. The comparisons between classes at each delay use the main branches of the delay tree.

***Silver-flower and virus data***

The silver-flower patterns of DPR's study[30], recorded at PAL-XFEL with a multi-port CCD detector and a photon energy of 5 keV, were used without the centrosymmetric filling of the missing region applied in that study. The PBCV-1 (r0227) and T4 (r0089) patterns from the Coherent X-ray Imaging Data Bank (IDs 14 and 13)[49,50], recorded at LCLS with a pnCCD detector and a photon energy of 1.2 keV, were converted to photons and masked as in DPR's study. Of the 475 hits with at least $10^4$ photons, the 310 with virus signals (124 PBCV-1 and 186 T4) were used for the comparisons. All methods were run as for the nanorod data.

***Optimized FP16 engines and throughput measurement***

Because NVIDIA TensorRT supports neither complex tensors nor Fourier transforms, the transforms between *r*- and *k*-space were written as exact real-valued matrix products for deployment, and the spectral branches of the networks were replaced by custom CUDA kernels, run as TensorRT plugin layers, that compute real FFTs in FP16 fused with the branch convolutions. In the resulting engines, the networks run in FP16, whereas the input processing, the transforms between *r*- and *k*-space and the amplitude blending are in FP32; each engine is built for its best batch size and replayed as a CUDA graph (optimized FP16 engine). Throughput was measured with PyTorch 2.14, TensorRT 11.4 and CUDA 12.6 on one NVIDIA RTX 6000 Ada GPU for batch sizes from 1 to 64 and is reported at each model's best batch size. Apart from the optimized engines, each network ran in its training precision, and HIO and GPS ran their 100 initial phases in parallel on the same GPU. FLOPs (2 × multiply–accumulate operations) are given per pattern at batch size 1.

***Significance tests and confidence intervals***

Differences between k2rNN and each other method, which were given the same patterns, were tested with the Wilcoxon signed-rank test, and differences between all pairs of rod classes at each delay, which are independent samples, with the Mann–Whitney U test; all tests were two-sided and not corrected for multiple comparisons. In the figures, * denotes $P < 0.05$, ** $P < 0.01$, *** $P < 0.001$, **** $P < 10^{-4}$ and ns $P \geq 0.05$. Bootstrap confidence intervals were computed from 4,000 resamples.

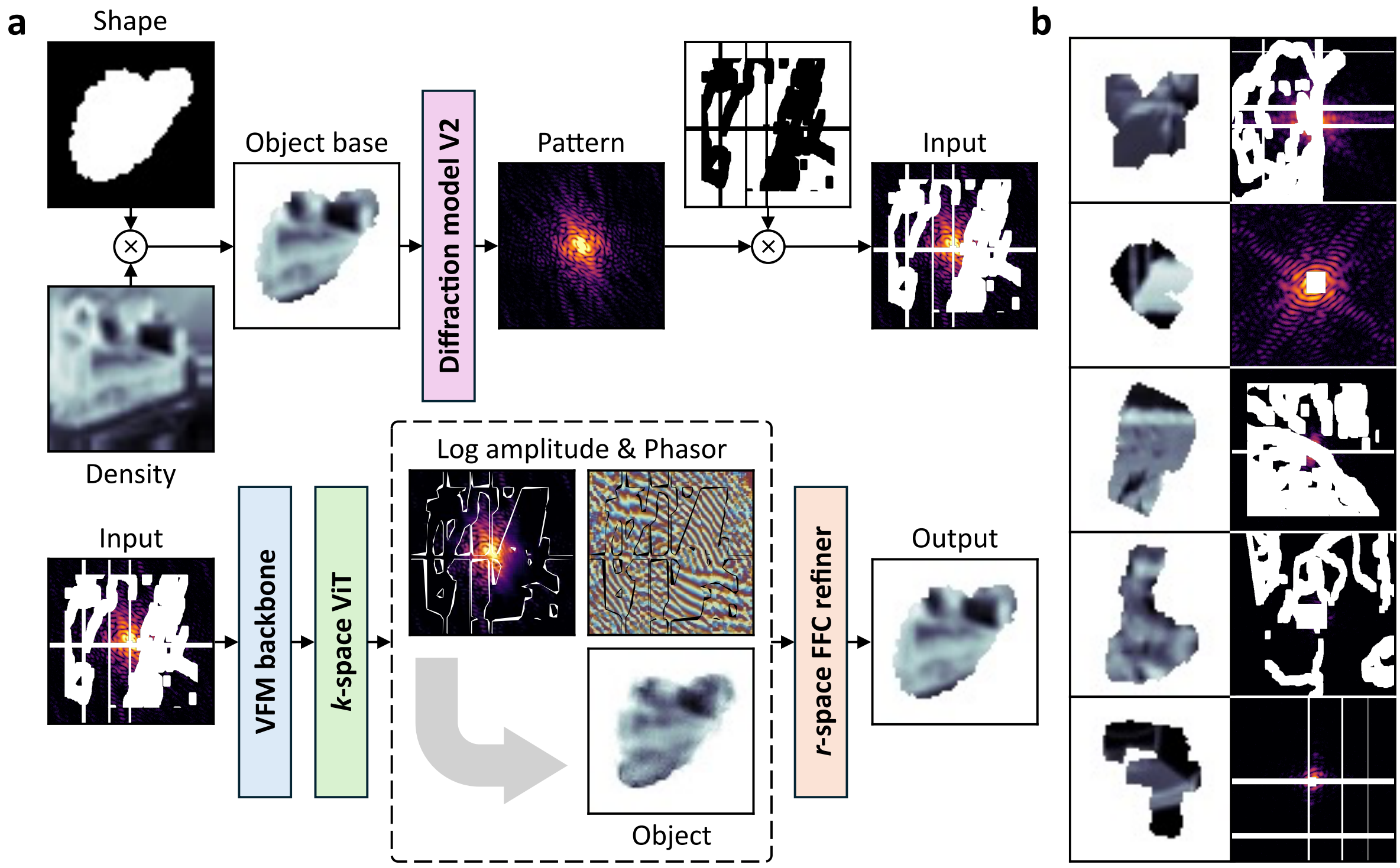

**Figure 1. Diffraction model V2 and k2rNN. a**, Diffraction model V2 (top) and k2rNN (bottom). Top: a binary shape (an EMNIST character) multiplied by a density (a CIFAR-100 crop) forms the object, which Diffraction model V2 turns into a diffraction pattern applying Ewald-sphere curvature, complex refractive index, partial coherence, background and detector effects; masks of the beamstop, detector gaps and irregular missing regions (white) complete the network input. Bottom: the VFM backbone encodes the input; the *k*-space ViT predicts the complex field as log amplitude and phasor (shown with the outlines of the masked regions); an inverse Fourier transform gives an initial *r*-space image, which the *r*-space FFC refiner turns into the final output image. The architecture of k2rNN is detailed in Supplementary Fig. 1. **b**, Further examples of objects (left) and input patterns (right). Patterns are shown on a log scale.

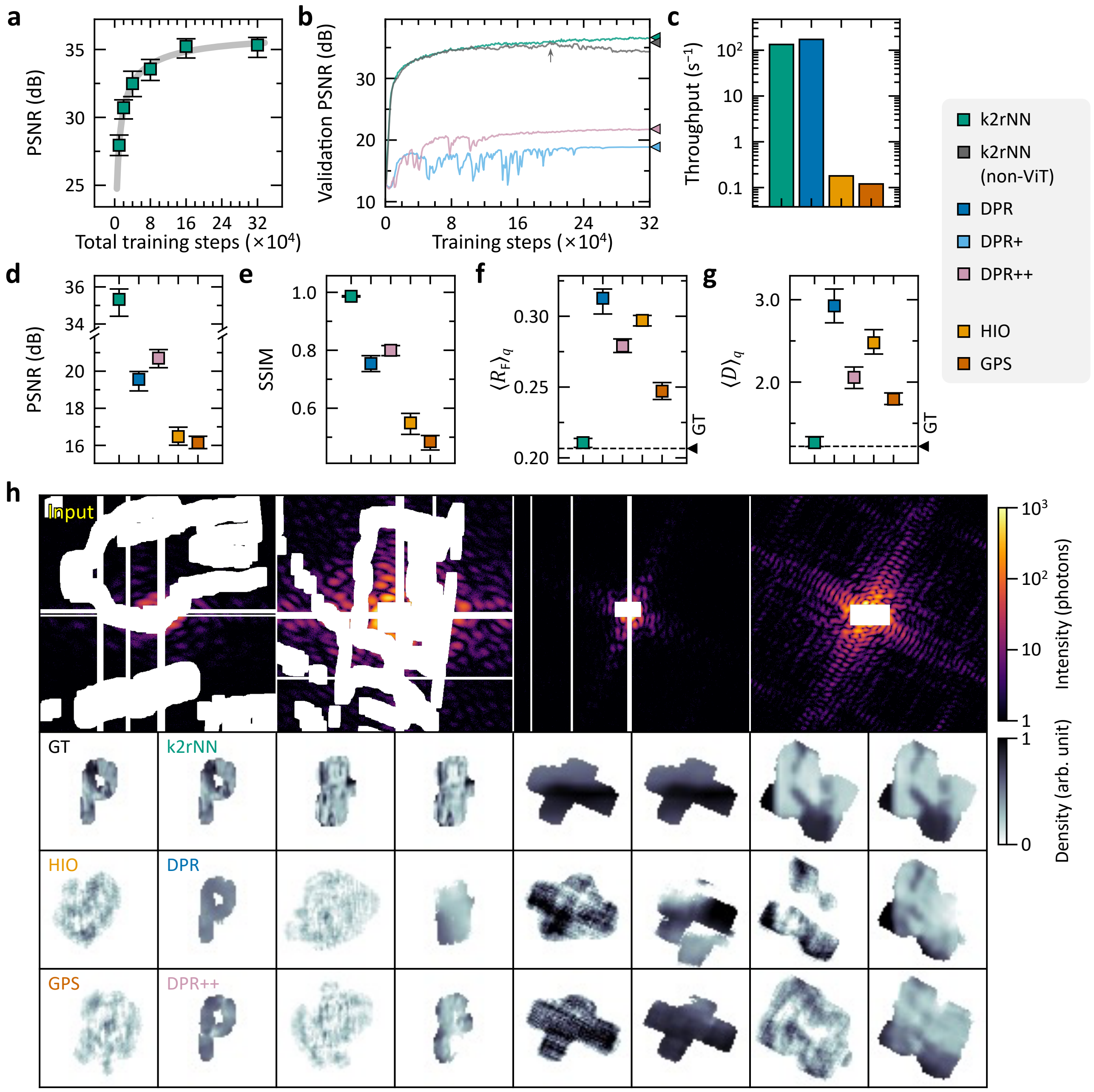

**Figure 2. Performance on simulated patterns. a**, k2rNN's test PSNR on 1,000 test patterns versus the number of training steps. The PSNR saturates with increasing training steps. **b**, Validation PSNR during training of k2rNN, the non-ViT variant of k2rNN and DPR retrained on Diffraction model V2 with DPR's own loss (DPR+) or ours (DPR++); triangles indicate each maximum, the arrow the maximum of k2rNN (non-ViT). **c**, Throughput on one NVIDIA RTX 6000 Ada GPU at each method's best batch size. **d**–**g**, PSNR (**d**), SSIM (**e**), $\langle R_F \rangle_q$ (**f**) and $\langle D \rangle_q$ (**g**) of k2rNN, DPR, DPR++, HIO and GPS on 1,000 test patterns. Dashed lines (GT), the medians of the ground-truth objects. k2rNN differs

from every other method in every metric ($P < 10^{-4}$). **h**, Representative examples. Input patterns (top) and the ground truth with their reconstructions by k2rNN, HIO, DPR, GPS and DPR++ (bottom; aligned and scaled to the ground truth) are shown. In **a** and **d**–**g**, markers indicate the median and bootstrap 95% confidence interval.

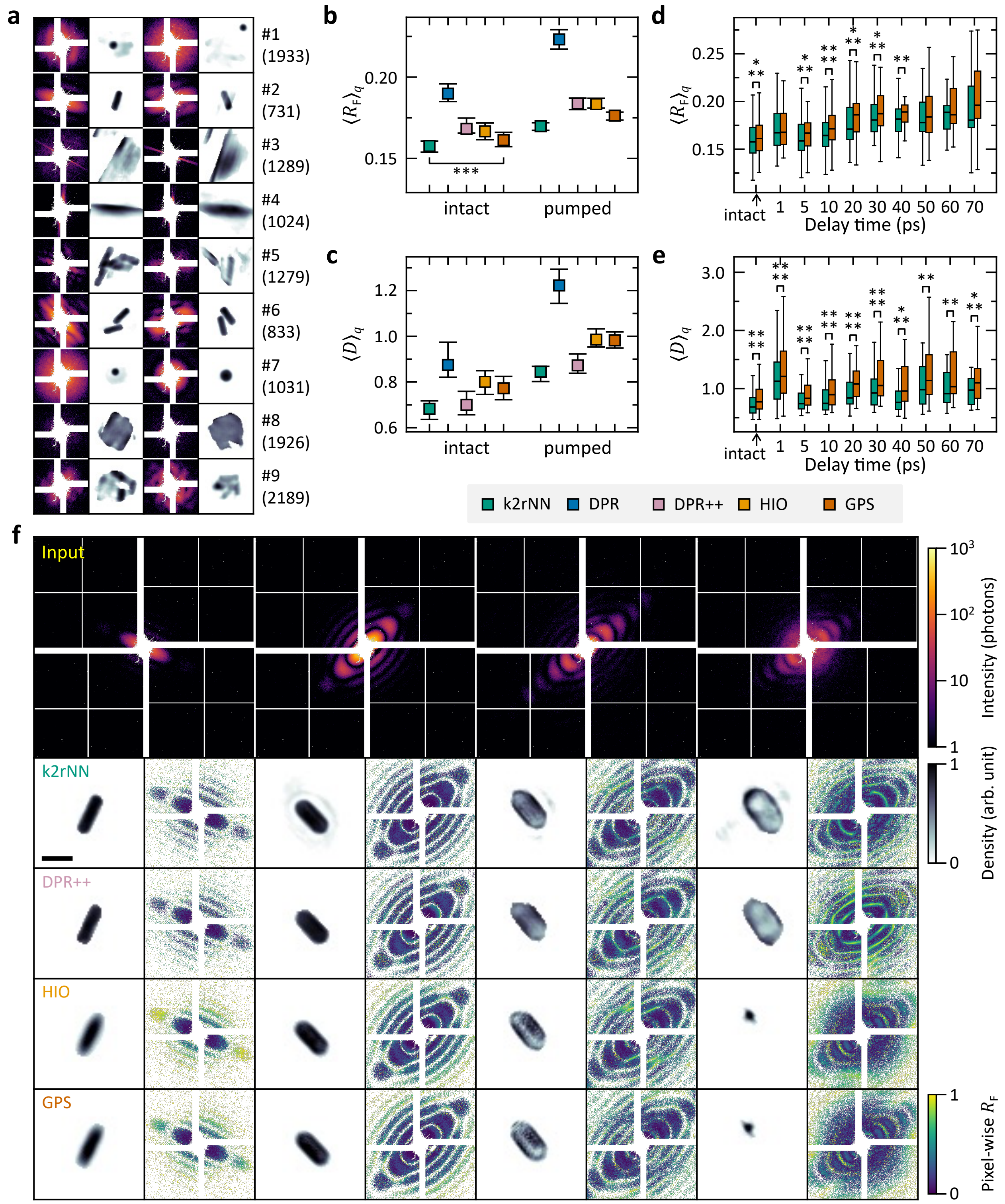

**Figure 3. Application to a real pump–probe experiment. a**, The nine groups from *k*-means clustering of the embeddings of 12,235 k2rNN reconstructions by the vision encoder of Qwen3.5-9B (hits per group in parentheses), each with two representative patterns and reconstructions. **b**,**c**, $\langle R_F \rangle_q$

(**b**) and $\langle D\rangle_q$ (**c**) of k2rNN, DPR, DPR++, HIO and GPS on 144 intact (no pump or $\Delta t \leq 0$) and 550 pumped ($0 < \Delta t \leq 70$ ps) single rods. Brackets indicate the significance of k2rNN versus each method (pairs with $P < 10^{-4}$ not marked). **d**,**e**, $\langle R_F\rangle_q$ (**d**) and $\langle D\rangle_q$ (**e**) of k2rNN and GPS versus the delay ($n$ = 144, 67, 88, 171, 61, 44, 35, 37, 17 and 30 from intact to 70 ps). Box plots present the median, quartiles and whiskers to 1.5 times the interquartile range without outliers, and brackets indicate all delays with significant differences. **f**, Representative examples of diagonal rods, intact and at 20, 50 and 70 ps. Input patterns (top), their reconstructions by k2rNN, DPR++, HIO and GPS (left; aligned and scaled to k2rNN's) and the corresponding pixel-wise $R_F$ (right; pixels with more than 0.5 photons) are shown. The scale bar is 100 nm. In **b**,**c**, markers indicate the median and bootstrap 95% confidence interval.

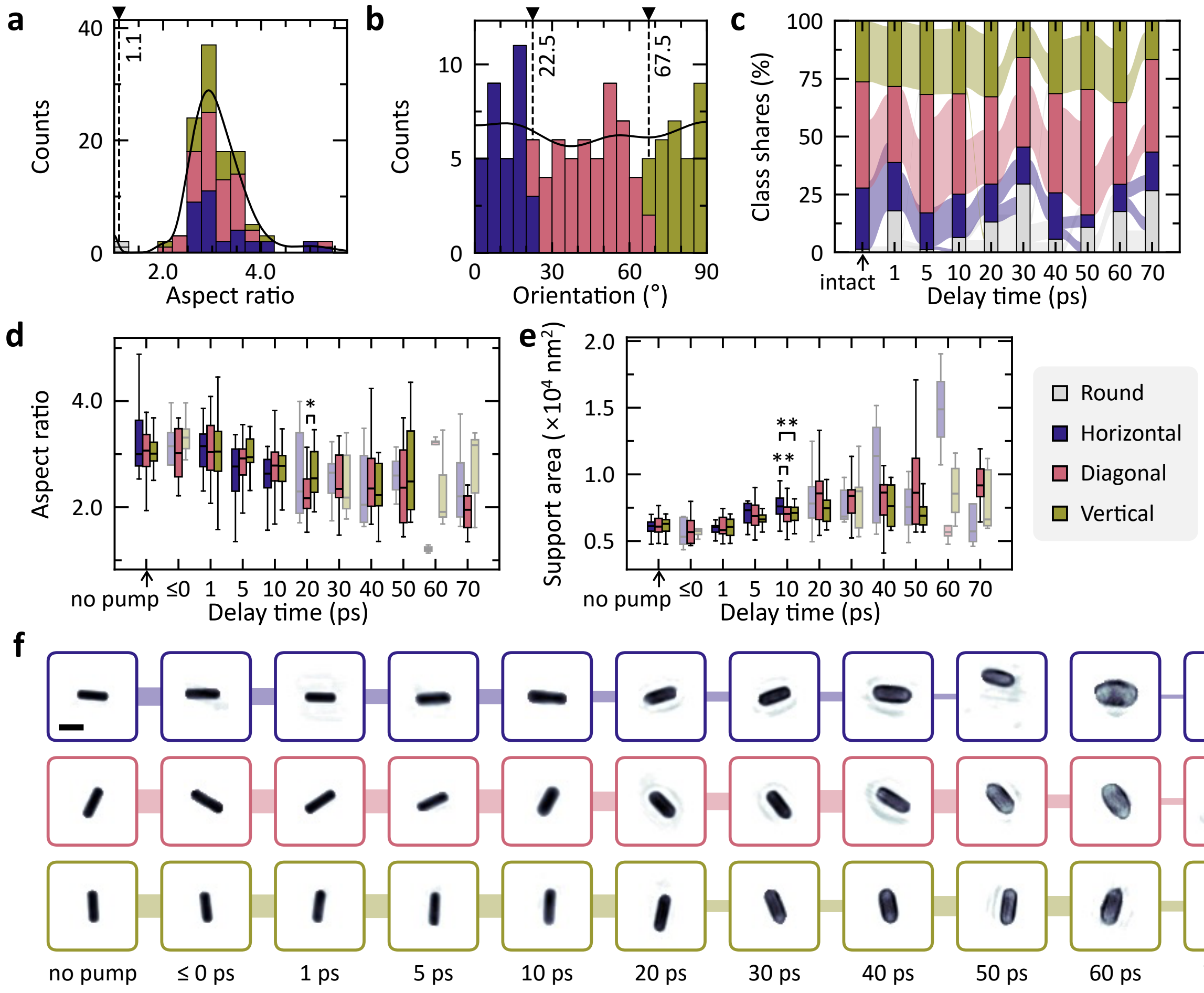

**Figure 4. Deformation of single gold nanorods across delays. a**,**b**, Aspect ratio (**a**) and long-axis angle from the horizontal (**b**) of the unpumped rods, stacked by class, with kernel density estimates (black lines). Dashed lines indicate the class boundaries at an aspect ratio of 1.1 and at 22.5° and 67.5°. **c**, Class shares versus the delay, with the flows between consecutive delays from unbalanced optimal transport. Intact comprises unpumped rods and rods pumped at $\Delta t \leq 0$. The flows couple independent populations and do not track individual rods. **d**,**e**, Aspect ratio (**d**) and support area (**e**) of the elongated rods in the main branches of the delay tree versus the delay, by class, with the unpumped rods (no pump) and the rods pumped at $\Delta t \leq 0$ (≤0) shown separately. Box plots present the median, quartiles and whiskers to 1.5 times the interquartile range without outliers ($n$ = 2–71 rods per box). Pale boxes contain fewer than 10 rods, and brackets indicate significant differences between classes at each delay. **f**, The delay tree, with one representative k2rNN reconstruction per node, linked between consecutive

delays by links whose widths are proportional to the fraction of rods transported. The scale bar is 100 nm.

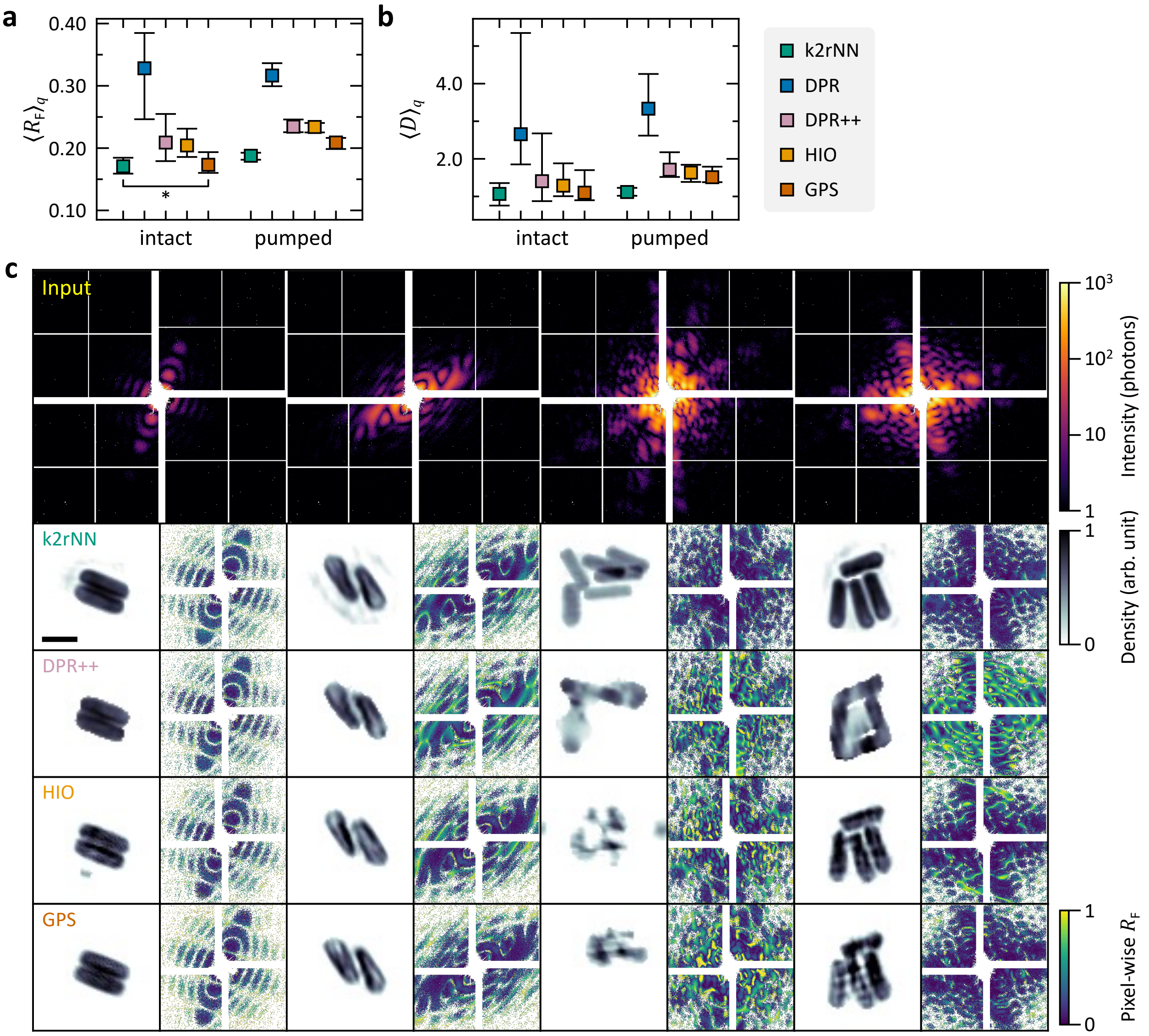

**Figure 5. Performance on multiple particles.** **a**,**b**, $\langle R_F \rangle_q$ (**a**) and $\langle D \rangle_q$ (**b**) of k2rNN, DPR, DPR++, HIO and GPS on 42 intact (no pump or $\Delta t \leq 0$) and 327 pumped ($0 < \Delta t \leq 70$ ps) multi-particle hits. Brackets indicate the significance of k2rNN versus each method (pairs with $P < 10^{-4}$ not marked). **c**, Representative examples of pumped multi-particle hits, rod pairs at 20 and 50 ps and rod clusters at 1 and 20 ps. Input patterns (top), their reconstructions by k2rNN, DPR++, HIO and GPS (left) and the corresponding pixel-wise $R_F$ (right) are shown. The scale bar is 100 nm. In **a**,**b**, markers indicate the median and bootstrap 95% confidence interval.

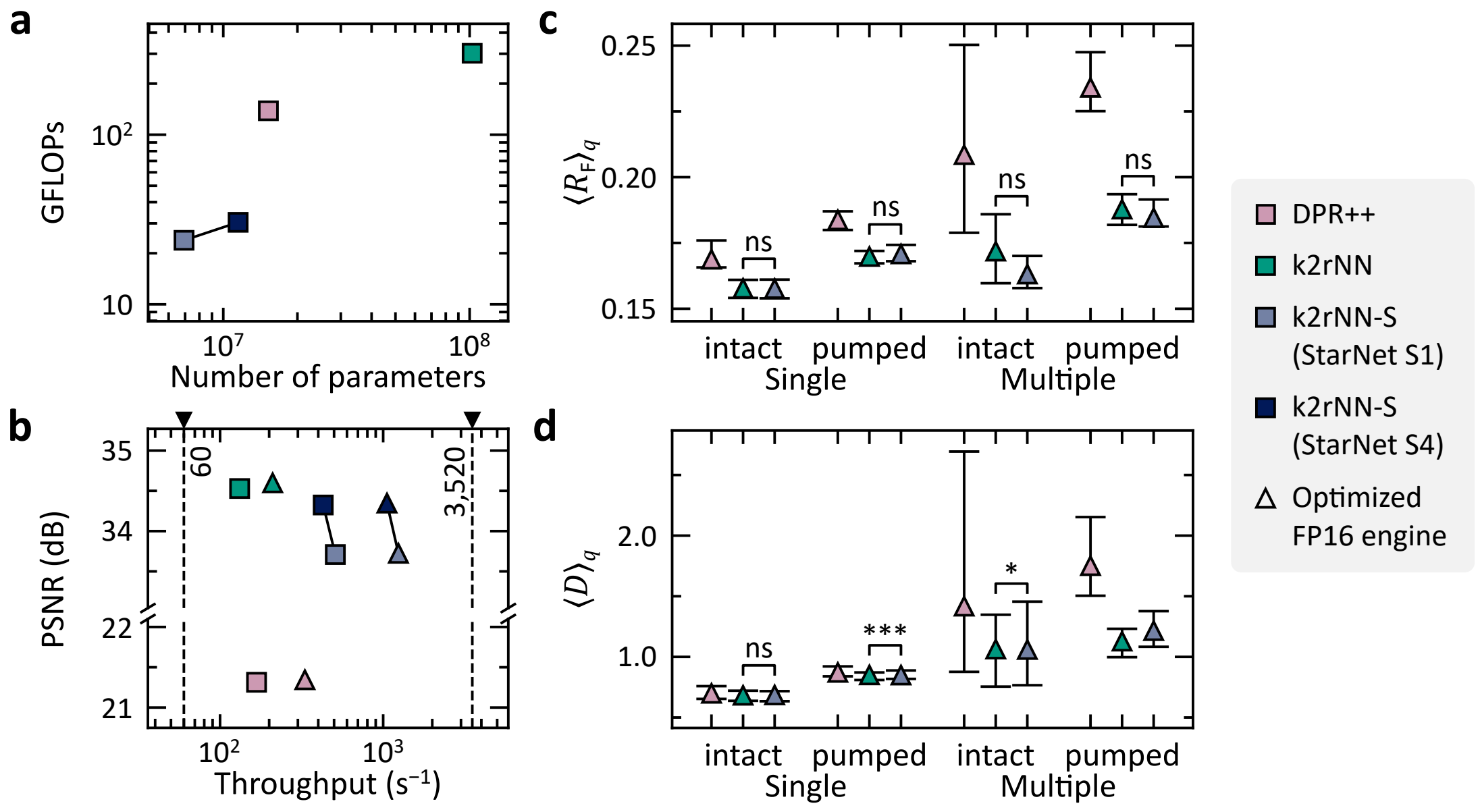

**Figure 6. Lightweight models and real-time deployment.** **a**, Compute versus the number of parameters of DPR++, k2rNN and k2rNN-S with StarNet S1 and S4 backbones. **b**, Test PSNR (mean over 1,000 simulated patterns) versus throughput at the best batch size on one NVIDIA RTX 6000 Ada GPU, in training precision (squares) and as optimized FP16 engines (triangles). Dashed lines indicate the repetition rate of PAL-XFEL (60 Hz) and the frame rate of the AGIPD detector at European XFEL (3,520 frames per second). **c**,**d**, $\langle R_F \rangle_q$ (**c**) and $\langle D \rangle_q$ (**d**) of the optimized FP16 engines of DPR++, k2rNN and k2rNN-S (StarNet S1) on the 1,063 labelled hits up to 70 ps, split into intact and pumped single rods and multi-particle hits. Brackets indicate the significance of k2rNN versus each engine (pairs with $P < 10^{-4}$ not marked). In **c**,**d**, markers indicate the median and bootstrap 95% confidence interval.

## DATA AVAILABILITY

The nanorod and silver-flower data are available from the corresponding authors on request, and the virus data from the Coherent X-ray Imaging Data Bank (IDs 13 and 14)[50].

## CODE AVAILABILITY

The source code and trained weights of all models and the graphical application CDI Workbench are available at https://github.com/sungyun98/k2rNN.

## ASSOCIATED CONTENTS

### Supplementary Information

- Supplementary Information (PDF)

## AUTHOR INFORMATION

### Corresponding Author

**Changyong Song** – *Department of Physics, POSTECH, Pohang 37673, Republic of Korea; Photon Science Center, POSTECH, Pohang 37673, Republic of Korea; Center for Ultrafast Science on Quantum Matter, Max Planck POSTECH/Korea Research Initiative, Pohang 37673, Republic of Korea; E-mail: cysong@postech.ac.kr*

### Authors

**Sung Yun Lee** – *Department of Physics, POSTECH, Pohang 37673, Republic of Korea; Photon Science Center, POSTECH, Pohang 37673, Republic of Korea; Center for Ultrafast Science on*

*Quantum Matter, Max Planck POSTECH/Korea Research Initiative, Pohang 37673, Republic of Korea*

**Eunyoung Park** – *Department of Physics, POSTECH, Pohang 37673, Republic of Korea; Photon Science Center, POSTECH, Pohang 37673, Republic of Korea; Center for Ultrafast Science on Quantum Matter, Max Planck POSTECH/Korea Research Initiative, Pohang 37673, Republic of Korea*

**Sinwoo Kim** – *Department of Physics, POSTECH, Pohang 37673, Republic of Korea; Photon Science Center, POSTECH, Pohang 37673, Republic of Korea; Center for Ultrafast Science on Quantum Matter, Max Planck POSTECH/Korea Research Initiative, Pohang 37673, Republic of Korea*

**Jungchan Choi** – *Department of Physics, POSTECH, Pohang 37673, Republic of Korea; Photon Science Center, POSTECH, Pohang 37673, Republic of Korea; Center for Ultrafast Science on Quantum Matter, Max Planck POSTECH/Korea Research Initiative, Pohang 37673, Republic of Korea*

**Seung-Phil Heo** – *Department of Physics, POSTECH, Pohang 37673, Republic of Korea; Photon Science Center, POSTECH, Pohang 37673, Republic of Korea; Center for Ultrafast Science on Quantum Matter, Max Planck POSTECH/Korea Research Initiative, Pohang 37673, Republic of Korea*

**Junha Hwang** – *Photon Science Center, POSTECH, Pohang 37673, Republic of Korea; XFEL Beamline Division, Pohang Accelerator Laboratory, POSTECH, Pohang 37673, Republic of Korea*

**Do Geun Jang** – *XFEL Beamline Division, Pohang Accelerator Laboratory, POSTECH, Pohang 37673, Republic of Korea*

**Daewoong Nam** – *Photon Science Center, POSTECH, Pohang 37673, Republic of Korea; XFEL Beamline Division, Pohang Accelerator Laboratory, POSTECH, Pohang 37673, Republic of Korea*

**Sangsoo Kim** – *XFEL Beamline Division, Pohang Accelerator Laboratory, POSTECH, Pohang 37673, Republic of Korea*

**Author contributions**

C.S. conceived the project. S.Y.L. developed Diffraction model V2, k2rNN and CDI Workbench and carried out the training, data processing and analyses. All authors took part in the experiment at PAL-XFEL. S.Y.L. and C.S. wrote the manuscript.

**Competing interests**

The authors declare no competing interests.

## ACKNOWLEDGEMENTS

This work was supported by the National Research Foundation of Korea (Grant Nos. RS-2022-NR068223 and RS-2024-00346711). S.Y.L. acknowledges support from the National Research Foundation of Korea (Grant No. RS-2026-25564880). The experiment at the NCI beamline of PAL-XFEL, Pohang Accelerator Laboratory, was approved by the Korean Synchrotron User's Association (Proposal No. 2024-2nd-NCI-014) and funded by the Ministry of Science and ICT of Korea. S.Y.L. thanks Jaechang Kim (POSTECH), Seokju Chung (Columbia University) and Changgeun Ji (ZETIC.ai) for helpful discussions on AI. The authors used Claude (Anthropic) to assist with language editing and code refactoring.

# Supplementary Information for

## Accurate non-iterative phase retrieval for time-resolved coherent diffraction imaging at X-ray free electron lasers

Sung Yun Lee[1,2,3], Eunyoung Park[1,2,3], Sinwoo Kim[1,2,3], Jungchan Choi[1,2,3], Seung-Phil Heo[1,2,3], Junha Hwang[2,4], Do Geun Jang[4], Daewoong Nam[2,4], Sangsoo Kim[4], and Changyong Song[1,2,3,*]

*[1]Department of Physics, POSTECH, Pohang 37673, Republic of Korea; [2]Photon Science Center, POSTECH, Pohang 37673, Republic of Korea; [3]Center for Ultrafast Science on Quantum Matter, Max Planck POSTECH/Korea Research Initiative, Pohang 37673, Republic of Korea; [4]XFEL Beamline Division, Pohang Accelerator Laboratory, POSTECH, Pohang 37673, Republic of Korea; [*]Corresponding author, e-mail: cysong@postech.ac.kr*

**CONTENTS**

**Supplementary Note 1. Ewald-sphere curvature in Diffraction model V2**

*Column approximation.* An object of normalized projected thickness $s(x, y)$ is $t = sD$ thick along the beam, where $D$ is its maximum thickness. Each column $(x, y)$ carries the exit wave $o(x, y)$ of the object (Methods), and its scattering is spread uniformly over the column's depth. In units of $D$, the column extends from $a = (\alpha - 1)s/2$ to $b = (\alpha + 1)s/2$, so that its length is $s$ and its mid-plane lies at $\zeta_0 = \alpha s/2$, with $\alpha$ drawn uniformly from −1 to 1 for each object.

*Far field.* On the Ewald sphere, the scattering vector at the detector has a longitudinal component $q_z = k(1 - \cos 2\theta) \approx |\boldsymbol{q}_\perp|^2/(2k)$, so scattering from depth $\zeta D$ acquires the phase $\exp(-\mathrm{i}q_z\zeta D)$. With $q_zD = \kappa\rho^2$, where $\rho = |\boldsymbol{q}_\perp - \boldsymbol{q}_0|/q_{\max}$, $\boldsymbol{q}_0$ is the beam centre, offset from the centre of the pattern by the beam-centre error $\boldsymbol{\Delta}$, and $q_{\max}$ is the largest $|\boldsymbol{q}_\perp|$ on the pattern, the far field is

$$F(\boldsymbol{q}_\perp) = \int d\boldsymbol{x}\, \exp(-\mathrm{i}\boldsymbol{q}_\perp \cdot \boldsymbol{x})\, o(\boldsymbol{x}) \frac{1}{s(\boldsymbol{x})} \int_{a(\boldsymbol{x})}^{b(\boldsymbol{x})} d\zeta\, \exp(-\mathrm{i}\kappa\rho^2\zeta)\,. \quad (1)$$

The inner integral evaluates to $\exp(-\mathrm{i}\kappa\rho^2\zeta_0)\,\mathrm{sinc}(\kappa\rho^2 s/2)$, with $\mathrm{sinc}(u) = \sin(u)/u$, the shape transform of the column along the beam. It depends on both $\boldsymbol{q}_\perp$, through $\rho$, and $\boldsymbol{x}$, through $s$ and $\zeta_0$, so it cannot be applied as a factor to the pattern, and equation (1) is not a single Fourier transform; evaluating it directly would need one Fourier transform per depth slice. Expanding $\exp(-\mathrm{i}\kappa\rho^2\zeta)$ in powers of $\zeta$ separates the two dependences:

$$F(\boldsymbol{q}_\perp) = \sum_n \frac{(-\mathrm{i}\kappa\rho^2)^n}{n!} \mathrm{FT}[om_n](\boldsymbol{q}_\perp)\,, \quad m_n = \frac{b^{n+1} - a^{n+1}}{(n+1)s}\,, \quad (2)$$

where $m_n$ is the column average of $\zeta_n$, the $n$-th depth moment of the column. In terms of $s$ and $\alpha$, $m_n = s^n[(\alpha + 1)^{n+1} - (\alpha - 1)^{n+1}]/[2^{n+1}(n + 1)]$; thus $m_0 = 1$ gives the projection and the moments with $n > 0$ make the pattern deviate from the projection. Because $|m_n| \leq 1$ and $\rho \leq 1$ (or slightly more with $|\boldsymbol{\Delta}| > 0$), $\kappa^n/n!$ sets the size of the $n$-th term, and the series keeps the terms $n = 0, \ldots, N - 1$, where $N$ is the first $n$ with $\kappa^n/n! < 10^{-3}$ ($N = 5$, 10 and 18 for $\kappa = 0.5$, 2 and 5). The beam-centre error $\boldsymbol{\Delta}$ is applied exactly as the phase ramp $R = \exp(2\pi i \boldsymbol{\Delta}\cdot\boldsymbol{x}/512)$ on $o$. For $\kappa \rightarrow 0$, only the $n = 0$ term remains ($m_0 = 1$), which is exactly the projection approximation used by DPR[1].

*Curvature strength*. At the corners of the pattern, $\kappa = q_z D = Dq_{\max}^2/(2k) \approx kD(2\theta_{\max})^2/2$, where $2\theta_{\max}$ is the largest scattering angle on the pattern. Because $\varphi_{\max} = k\delta D$, the ratio $\kappa/\varphi_{\max} = (2\theta_{\max})^2/(2\delta)$ depends only on the geometry and the material, so Diffraction model V2 draws $\kappa = c\varphi_{\max}$, with the ratio $c$ log-uniform between 0.2 and min(200, $5/\varphi_{\max}$) so that $\kappa \leq 5$. For the nanorod data ($2\theta_{\max} = 0.0173$ rad; $\delta = 3.74 \times 10^{-5}$ and $\beta = 3.23 \times 10^{-6}$ for gold at 9 keV[2]), $c = 4.0$, $\varphi_{\max} = 0.085$ and $\kappa = 0.34$ for a rod 50 nm thick, and $r = \beta/\delta = 0.086$; all four lie within the ranges of Diffraction model V2.

*Accuracy and validity*. On eight random weak-phase objects ($\varphi_{\max} = 0.01$) with $\kappa = 0.5$, 2 and 5, equation (2) agrees with a direct evaluation of equation (1) on 400 depth slices to a relative L2 difference in $F$ of at most $7 \times 10^{-7}$, $2 \times 10^{-6}$ and $5 \times 10^{-6}$, respectively, whereas the projection approximation deviates by up to 1.6%, 5.5% and 11%. This column approximation is the first Born approximation when the phase is weak; when the phase is strong, the column approximation neglects multiple scattering within the object. A flat detector at a distance $L$ from the sample captures $|\boldsymbol{q}_\perp| = k \sin[\arctan(r_d/L)]$ at a radial distance $r_d$ from the beam centre, rather than $kr_d/L$ as on the regular Fourier grid; in the nanorod geometry, this moves the corner of the pattern by 0.05 pixels and changes the solid angle per pixel by 0.05%, so both effects are neglected.

| # | Effect | Parameter, range (sampling) | Probability |
|---|---|---|---|
| 1 | Shape | EMNIST character (balanced split), binarized at 0.1; square dilation 3–7 px (integer) on the native 28 px | 1 |
| 2 | Orientation and size | Flip; rotation 0–360° (lin); isotropic scaling to the longest side of 20–60 px (lin); binarized after the transform | 1 |
| 3 | Density | CIFAR-100 image, greyscaled, randomly cropped, resized to shape extent; $s$ = shape × density, maximum 1, in 64 × 64 window | 1 |
| 4 | Material | Exit wave $o = \exp[(\mathrm{i} - r)\varphi_{\max}s] - 1$ (target $\varphi = \varphi_{\max}s$); $r = \beta/\delta$ $5 \times 10^{-4}$–0.5 (log); $\varphi_{\max}$ $10^{-3}$–2 rad (log) | 1 |
| 5 | Ewald-sphere curvature | $F = \Sigma_n (-\mathrm{i}\kappa\rho^2)^n/n!\ \mathrm{FT}[om_n]$ (terms while $\kappa n/n! \geq 10^{-3}$); FT on 64 × 64 zero-padded to 512 × 512; $c$ 0.2–min(200, $5/\varphi_{\max}$) (log); $\alpha$ −1–1 (lin) [Supplementary Note 1] | 1 |
| 6 | Beam-centre error | Exact phase ramp, ±2 px per axis (lin) | 1 |
| 7 | Partial coherence | Gaussian Schell model on the autocorrelation; $\sigma$ = 0.2–50 × object size (log) | 1 |
| 8 | Photon count | $10^5$–$10^7$ photons over 512 × 512 (log) | 1 |
| 9 | Radial background | Flat + $(q + 5)^{-p}$ mixture (centred on the true beam), flat weight 0–1 (lin), $p$ 2–4 (lin); level 0.001–1 × $S_{\mathrm{ref}}$ (log) | 0.2 |
| 10 | Streaks | 1–2 lines through the beam centre with a Gaussian cross-section of FWHM 1–10 px (lin), decaying as $(q + 5)^{-p}$ along the line, $p$ 2–4 (lin); level 0.001–1 × $S_{\mathrm{ref}}$ (log) | 0.1 |
| 11 | Shot noise | Poisson(signal + background) | 1 |
| 12 | Gain errors | Pixel 0.5–5% and block 1–10% on a 1–4 × 1–8 grid (lin) | 0.2 |
| 13 | Readout noise | Gaussian, 0.02–0.5 photon (log) | 1 |
| 14 | Offsets | Each with probability 0.6 (all log; in photons): block offsets 0.001–0.2, row and column banding 0.001–0.1, fixed-pattern noise 0.002–0.2, smooth offset field 0.001–0.1 with correlation length 5–200 px | 0.8 |
| 15 | Bad pixels | Fraction $10^{-5}$–$2 \times 10^{-3}$ (log); 75% add Uniform(0, 1) × (1–20, log) photons, 25% set to zero (unmasked) | 0.8 × 0.6 |
| 16 | Background subtraction | $g(1 + \varepsilon)B$ subtracted, with $g$ the gain and $B$ the background; ε −0.2–0.2 (lin) | whenever #9 or #10 applies |
| 17 | Irregular missing regions | NVIDIA irregular masks; random flips | 1 |
| 18 | Beamstop | Rectangle, half sizes 5–50 px per axis (lin); centre within ±min(10, half size) px of the true beam | 1 |
| 19 | Detector gaps | 1–3 full-length bands per direction (each direction with probability 0.75); width 1–20 px (lin); the first band within ±20 px of the beam | 0.5 |

**Supplementary Table 1. Parameters of Diffraction model V2.** Effects are listed in the order of application. Probability is the fraction of patterns on which an effect is applied; ranges are sampled uniformly (lin) or log-uniformly (log). $S_{ref}$ is the object's mean signal in the shell 224–256 pixels from the true beam centre.

| | **k2rNN** | **k2rNN-S** |
|---|---|---|
| Backbone (channels per stage, projected) | DINOv3 ConvNeXt-B (384) | StarNet S1 / S4 (96) |
| *k*-space ViT: tokens, width, blocks, heads | 4,096, 192, 8, 6 | 1,024, 128, 6, 4 |
| *k*-space ViT: window, registers, MLP ratio | 256 tokens (alternating), 16, 4 (2 for token-to-register cross-attention) | Same |
| *r*-space FFC refiner: stages × blocks, channels | 4 × 3, 96 | 6 × 3, 32 |
| Parameters / compute | 102.4 M / 302 GFLOPs | 6.9 M / 24 GFLOPs (S1); 11.5 M / 30 GFLOPs (S4) |
| Optimizer | AdamW, lr $5 \times 10^{-4}$ (backbone × 0.2), weight decay $10^{-4}$, gradient clipping at 1.0, 1,000 warm-up steps, cosine decay to 0, EMA 0.9995 | Same |
| Training | 320,000 steps × 160 patterns on 4 NVIDIA RTX 6000 Ada and 12 NVIDIA GeForce RTX 3090 | 320,000 steps × 160 patterns on 4 NVIDIA RTX 6000 Ada |

**Supplementary Table 2. Hyperparameters of k2rNN and k2rNN-S.** Same indicates the setting of k2rNN. GFLOPs are counted as 2 × multiply–accumulate operations per pattern at batch size 1.

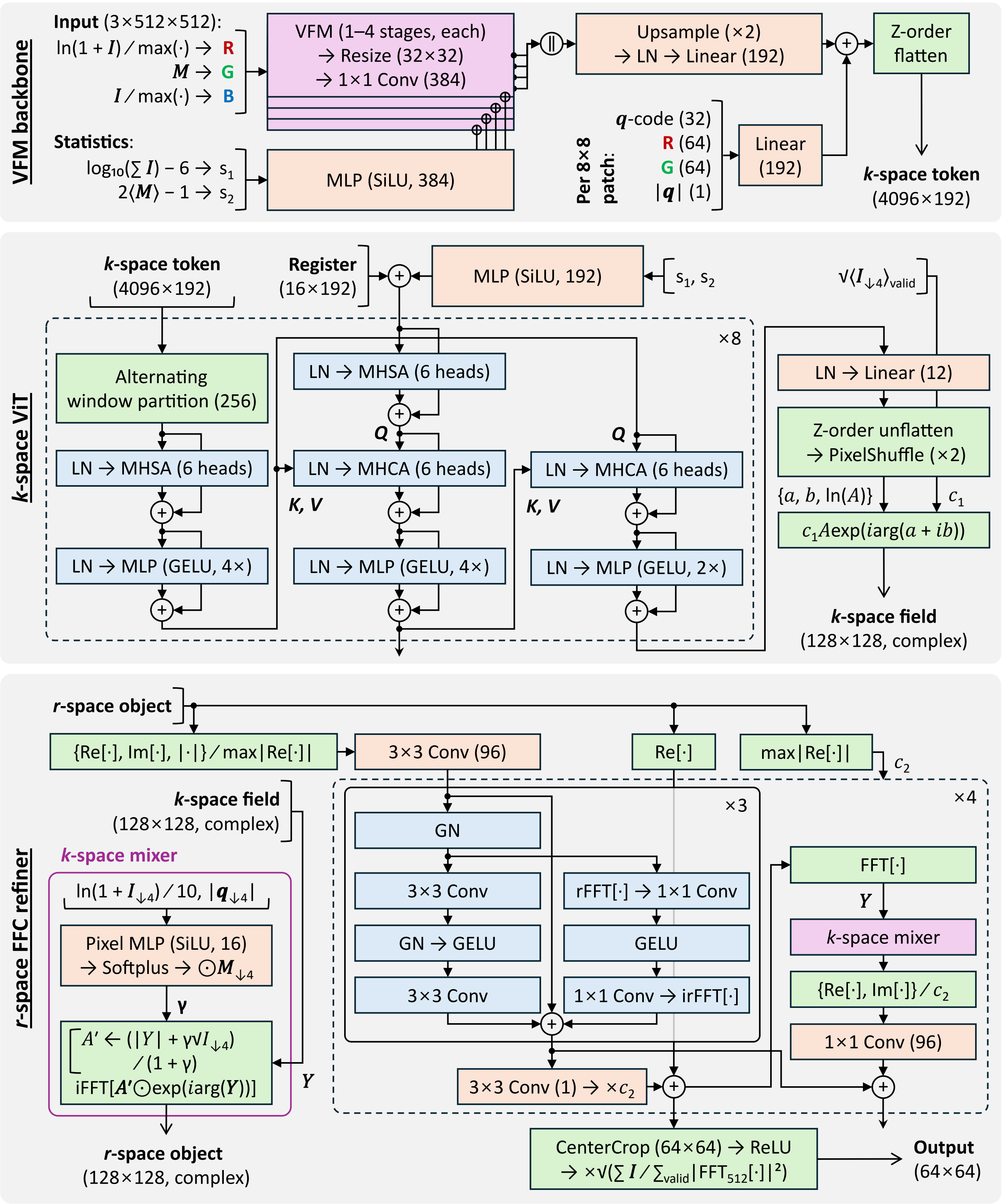

**Supplementary Fig. 1. Architecture of k2rNN. VFM backbone** (top). The input pattern $\boldsymbol{I}$ and its mask $\boldsymbol{M}$ are mapped to an RGB image as R = ln(1 + $\boldsymbol{I}$)/max(·), G = $\boldsymbol{M}$, B = $\boldsymbol{I}$/max(·). The features of the four stages of the VFM[3] are bilinearly resized to 32 × 32 and projected to 384 channels pixel-wise, and a SiLU MLP embedding of the statistics $s_1 = \log_{10}(\Sigma\boldsymbol{I}) - 6$ and $s_2 = 2\langle\boldsymbol{M}\rangle - 1$ is added to each stage's feature map. The concatenated features are bilinearly upsampled by a factor of 2,

normalized by layer normalization (LN), projected to 192 channels and added to the embedding of each 8 × 8 patch, also projected to 192 channels from the Fourier features of $\boldsymbol{q}$ ($\boldsymbol{q}$-code, 32), its R (64) and G (64) values and $|\boldsymbol{q}|$ (1). All 4,096 *k*-space tokens are flattened in Z-order[4,5]. ***k*-space ViT** (middle). The ViT consists of eight blocks of windowed multi-head self-attention (MHSA; alternating window partitions of 256 tokens[6]), register self-attention and register-to-token multi-head cross-attention (MHCA), and token-to-register cross-attention, each with LN, a GELU MLP and residual connections. The 16 registers[7] are learned embeddings plus a SiLU MLP embedding of $s_1$ and $s_2$. A final LN, 12-channel projection, Z-order unflattening and a pixel shuffle[8] by a factor of 2 give $a$, $b$ and $\ln A$ on the 128 × 128 grid, which form the *k*-space field $Y = c_1 A \exp(\mathrm{i} \arg(a + \mathrm{i}b))$, with $c_1 = \sqrt{\langle \boldsymbol{I}_{\downarrow 4} \rangle_{\mathrm{valid}}}$, the square root of the mean sampled intensity over the measured pixels. ***r*-space FFC refiner** (bottom). The *k*-space mixer blends the field $\boldsymbol{Y}$ with the measured amplitudes $\sqrt{\boldsymbol{I}_{\downarrow 4}}$ by a learned pixel-wise weight $\boldsymbol{\gamma}$, a softplus-activated pixel-wise SiLU MLP (Pixel MLP) of $\ln(1 + \boldsymbol{I}_{\downarrow 4})/10$ and $|\boldsymbol{q}_{\downarrow 4}|$ masked by $\boldsymbol{M}_{\downarrow 4}$, and returns the *r*-space object by an inverse FFT. Four stages of three FFC blocks adapted from FFC and LaMa[9,10] refine the object, normalized by $c_2 = \max|\mathrm{Re}[\cdot]|$. Each block has group normalization (GN), local 3 × 3 convolutions and a spectral branch (rFFT → 1 × 1 convolution → GELU → 1 × 1 convolution → inverse rFFT). Between stages, the updated object is Fourier transformed, passed through the *k*-space mixer and fed back through a 96-channel pixel-wise projection. The output is the central 64 × 64 window, rectified and scaled to the measured photon count.

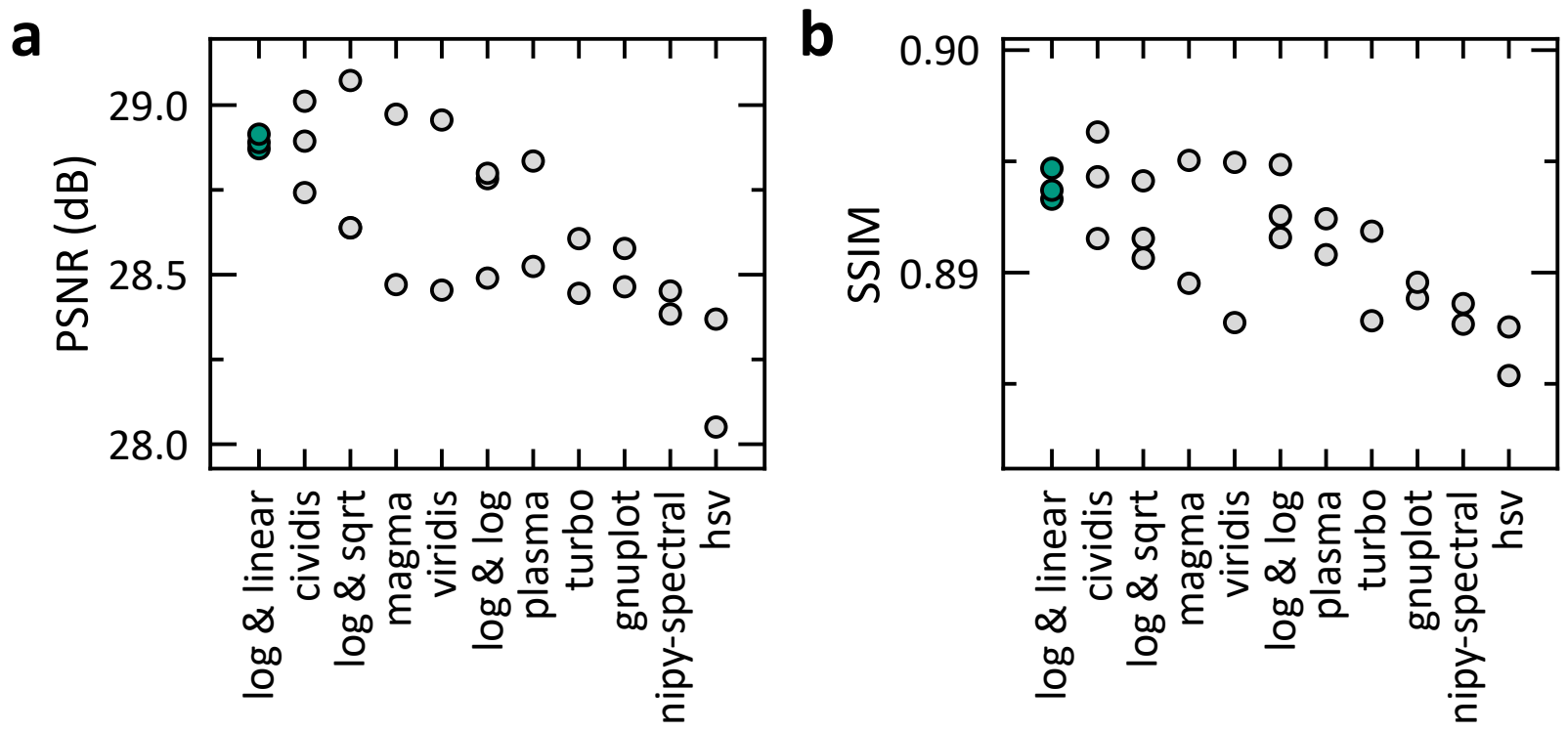

**Supplementary Fig. 2. Input colour mapping. a**,**b**, Test PSNR (**a**) and SSIM (**b**) on Diffraction model V2 of k2rNN pilot training runs (60,000 steps of 16 patterns) that differ only in how the pattern is mapped to the three input channels of the VFM backbone. Each marker indicates one training seed. The mappings are log & linear (ours, green), with R = $\ln(1 + \boldsymbol{I})/\max(\cdot)$, G = $\boldsymbol{M}$ and B = $\boldsymbol{I}/\max(\cdot)$; log & sqrt, with B = $\sqrt{(\boldsymbol{I}/\max(\cdot))}$; log & log, with B = R; and $\ln(1 + \boldsymbol{I})/\max(\cdot)$ mapped through the matplotlib colour maps cividis, magma, viridis, plasma, turbo, gnuplot, nipy-spectral and hsv. The mappings are ordered by mean PSNR. Log & linear (28.89 dB) and cividis (28.88 dB) are equal within the seed spread, and the cyclic colour map hsv performs worst (28.21 dB).

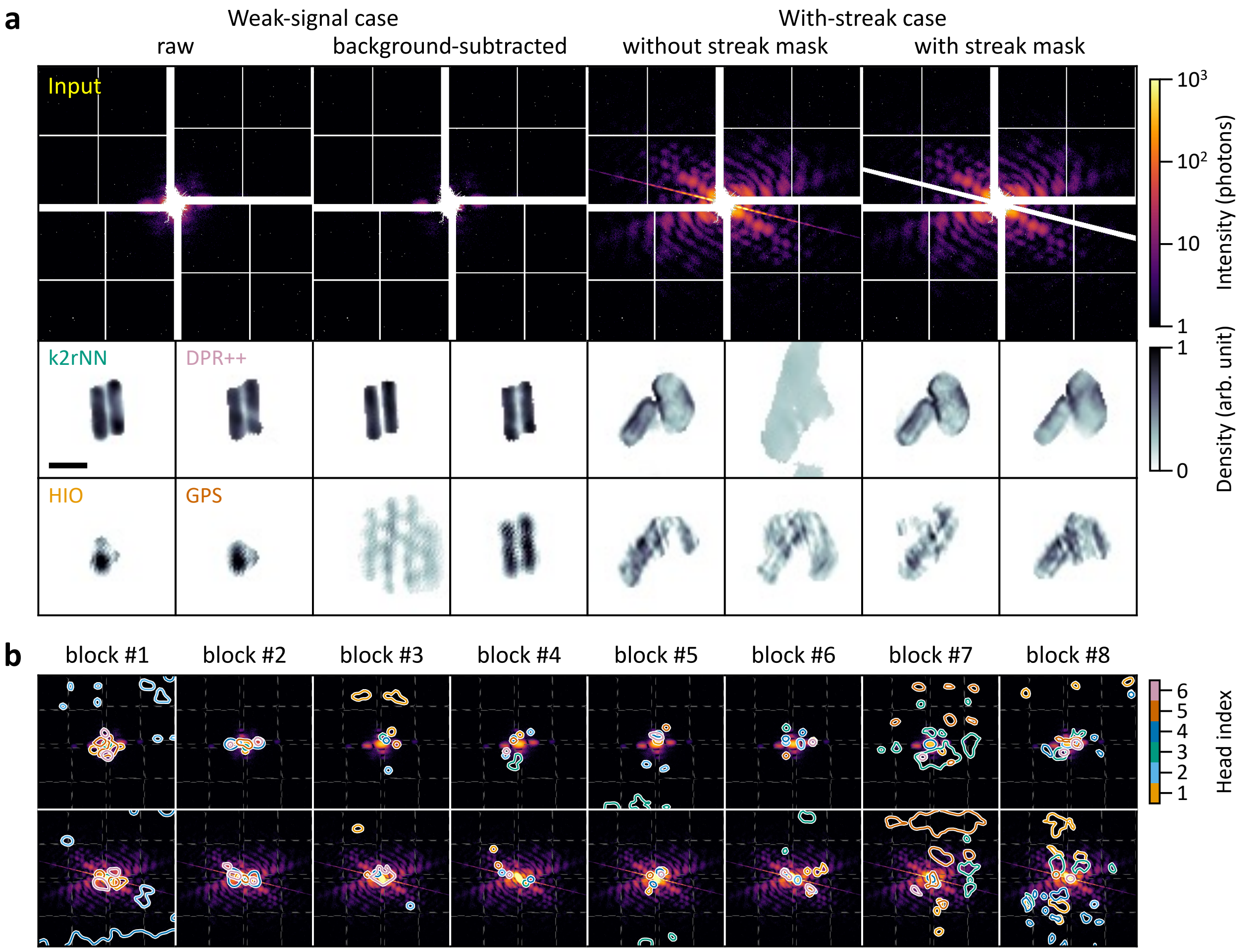

**Supplementary Fig. 3. Contaminated patterns and the attention of the *k*-space ViT. a**, Two measured nanorod patterns, each given as two inputs: a weak-signal pattern before (raw) and after background subtraction, and a pattern with a streak without and with a mask over the streak. Input patterns (top) and their reconstructions by k2rNN, DPR++, HIO and GPS (bottom) are shown. The scale bar is 100 nm. **b**, Attention of the 16 registers to the *k*-space tokens in blocks #1–#8 of the *k*-space ViT for the raw weak-signal pattern (top) and the unmasked with-streak pattern (bottom). Each head's attention map, averaged over the registers, is drawn as one contour at half its maximum in the colour of its head index. Behind the contours, the input pattern is shown on the measured pixels and k2rNN's predicted pattern, more translucent, on the unmeasured ones. Grey dashed lines indicate the edges of the unmeasured regions due to the detector-module gaps and the beamstop.

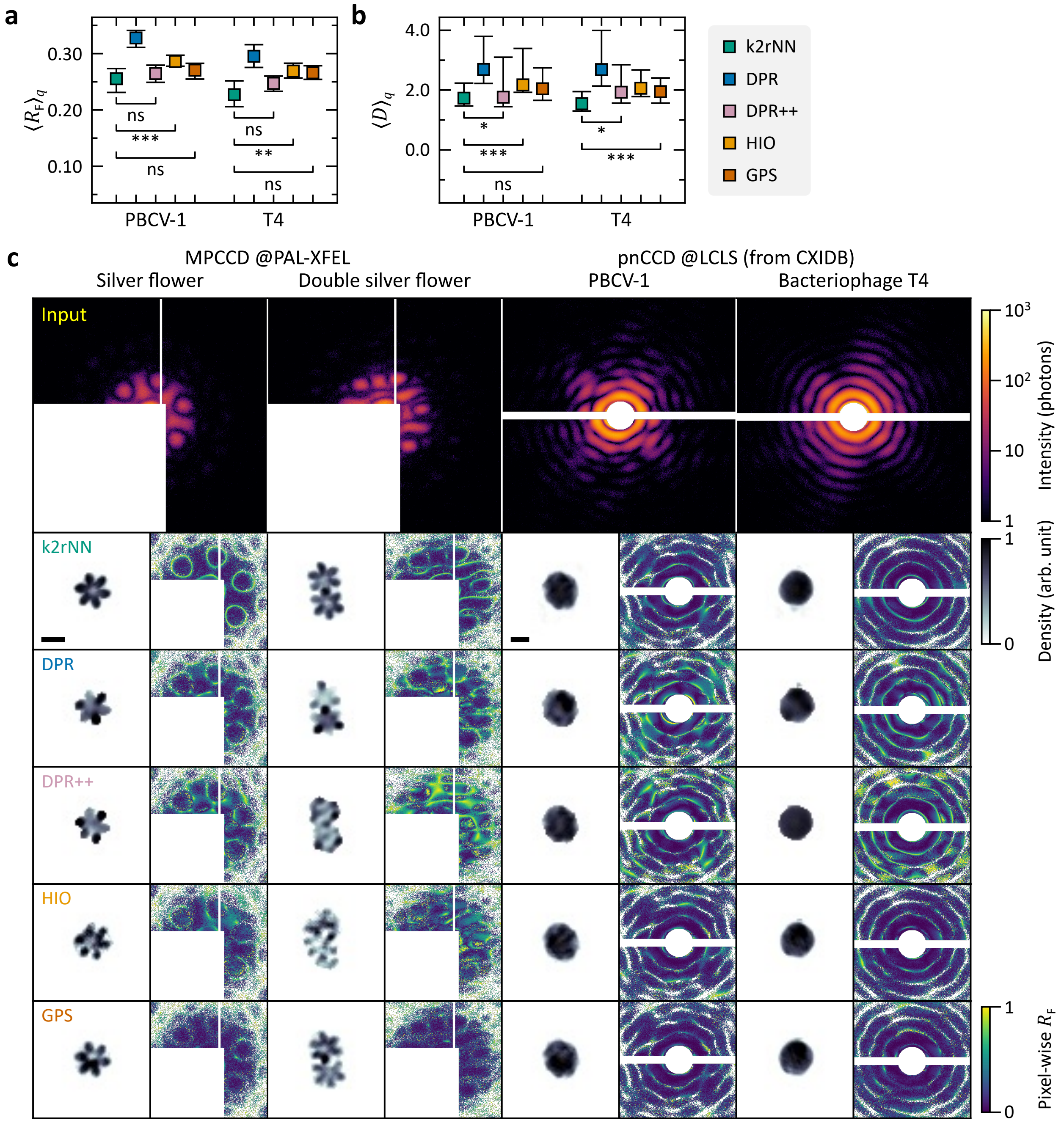

**Supplementary Fig. 4. Patterns from other instruments. a,b**, $\langle R_F \rangle_q$ (**a**) and $\langle D \rangle_q$ (**b**) of k2rNN, DPR, DPR++, HIO and GPS on 310 virus patterns from the Coherent X-ray Imaging Data Bank (124 PBCV-1 and 186 T4, unbinned). Brackets indicate the significance of k2rNN versus each method (pairs with $P < 10^{-4}$ not marked). **c**, Representative examples of a silver flower, a double silver flower, PBCV-1 and T4. The silver flowers were measured with a multi-port CCD detector at PAL-XFEL, and a quarter of each pattern was occluded by a beamstop. Input patterns (top), their reconstructions by k2rNN, DPR, DPR++, HIO and GPS (left) and the corresponding pixel-wise $R_F$ (right; the central 256 × 256 pixels) are shown. The $\langle R_F \rangle_q$ / $\langle D \rangle_q$ values of k2rNN, DPR, DPR++,

HIO and GPS are 0.154 / 1.03, 0.195 / 1.44, 0.236 / 2.05, 0.239 / 1.88 and 0.160 / 1.27 for the silver flower, and 0.170 / 1.40, 0.289 / 3.97, 0.357 / 5.21, 0.290 / 3.39 and 0.166 / 1.48 for the double silver flower. The scale bars are 200 nm. In **a**,**b**, markers indicate the median and bootstrap 95% confidence interval.